\documentclass[notoc,nohyper]{JHEP} 

\usepackage{epsfig}

\newcommand\fverb{\setbox\pippobox=\hbox\bgroup\verb}
\newcommand\fverbdo{\egroup\medskip\noindent%
			\fbox{\unhbox\pippobox}\ }
\newcommand\fverbit{\egroup\item[\fbox{\unhbox\pippobox}]}
\newbox\pippobox
\def\aves#1{{\langle #1 \rangle}}
\def\ssk{{\sss (k)}}
\def\ave#1{{\left\langle #1 \right\rangle}}
\def\sss{\scriptscriptstyle}
\def\sumxi{{\sum_j\xi^\ssk_j}}
\def\sfrac#1#2{{\textstyle{#1\over #2}}}
\def\NBR#1{{\left( #1 \right)}}
\def\acp{\alpha_{\sss CP}}
\def\scp{s_{\sss CP}}
\def\lsim{\;\raise0.3ex\hbox{$<$\kern-0.75em \raise-1.1ex\hbox{$\sim$}}\;}
\def\gsim{\;\raise0.3ex\hbox{$>$\kern-0.75em \raise-1.1ex\hbox{$\sim$}}\;}

\title{Erratum for\\ ``Supersymmetric Electroweak Baryogenesis''}

\author{James M.\ Cline\\
	McGill University, Montr\'eal, Qu\'ebec, Canada\\
	E-mail: \email{jcline@physics.mcgill.ca}}
\author{Michael Joyce\\
	LPT, Universit\'e Paris-XI, B\^atiment 211,
               F-91405 Orsay Cedex, France\\
       E-mail: \email{michael.joyce@th.u-psud.fr}}
\author{Kimmo Kainulainen\\
	NORDITA, Blegdamsvej 17, DK-2100, Copenhagen \O ,
          Denmark\\
	E-mail: \email{kainulai@nordita.dk}}

\abstract{We correct a numerical error which led to an overestimate of the
baryon asymmetry from supersymmetric electroweak baryogenesis in our paper 
JHEP07(2000)018.  Updated dependences of the baryon asymmetry on chargino 
mass parameters and the bubble wall velocity are shown. We also include 
LEP-II-constraints on the chargino mass parameters.  Combined with our 
corrected results for the baryon asymmetry these constraints imply that 
the phase in the chargino mass matrix must violate CP nearly maximally 
in order to generate a large enough baryon asymmetry.  A number of other 
typographical errors are also corrected.}

\begin{document} 

\maketitle 

\section{Correction of typographical errors}

The corrected version of several equations which had typographical
errors is given here.

\noindent Equation in footnote 2:
$\displaystyle
\dot {\bf p}_k = -\partial_{\bf x} H - e d_t {\bf A}
                       = e({\bf E} + {\bf v}\times {\bf B}).
$

\noindent Eq.\ (2.13):
$\displaystyle
\dot{p_c} = -(\partial_x \omega)_{p_c}=v_g \acp'
-\frac{|m||m|'}{(\omega + \scp \frac{s \theta'}{2})}
   + \scp\frac{s\theta''}{2}
$

\noindent Eq.\ (2.16):
$\displaystyle
(\partial_{p_c} v_g )_{x}=
\frac{|m^2|}{(\omega + \scp \frac{s \theta'}{2})^{3}}$\\
$\displaystyle
\phantom{....;}\qquad\qquad(\partial_x v_g )_{p_c} =
-\acp'
\frac{|m^2|}{(\omega + \scp \frac{s \theta'}{2})^{3}} -
v_g \frac{|m||m|'}{(\omega + \scp \frac{s \theta'}{2})^2}$

\noindent Eq.\ (2.23):
$\displaystyle
j^\mu(x) = \int \frac{p_{||} dp_{||}d\omega }{4\pi^2}
             \left(\frac{1}{v_g} \, ; \,\hat {\bf p}\right)
             f(\omega ).\nonumber   
$

\noindent Note: the preceding equation refers to the contributions of
particles
or antiparticles separately; thus the total current is the sum of both
contributions.

\medskip
\noindent Eq.\ (4.18):
$\displaystyle
      - \frac{\aves{v^2_{p_z}}}{\Gamma^t_i}\xi_i''
            - v_w \xi_i'
            - \sum_j \frac{\Gamma^d_{ik}}{\Gamma^t_i} {\xi^{(k)}_j} '
            + \Gamma^d_{ik} \sumxi
            =  -\frac{v_w \beta}{\Gamma^t_i} \ave{v_{p_z}\delta\! F_i}'.
$

\medskip
\noindent Eq.\ (4.23):
$\displaystyle
S_i \equiv - \kappa_i \frac{v_w D_i}{\aves{v^2_{p_z}} T}
                              \aves{v_{p_z}\delta\! F_i}'.
$

\medskip
\noindent Eq.\ (5.9):
$\displaystyle
       \Gamma_{y} \leftrightarrow       y h_2 \bar u_R q_L +
          y \bar u_R \tilde h_{2L}\tilde q_L
             + y \tilde u^*_R \tilde h_{2L} q_L 
$

\medskip
\noindent Eq.\ (5.41):
$\displaystyle
\xi_-(z) = - \sfrac{1}{36} R \frac{\Gamma_m}{\alpha_- \gamma D_h}
      e^{-(v_w/D_h)z} \, \int_{-\infty}^\infty dy\,  {\cal G}_-(y)
        \, S_H(y),
\qquad  z>0
$

\medskip
\noindent Eq.\ (5.54):
$\displaystyle
S_H = - \frac{\lambda}{2}\frac{v_w D_h}{\aves{v_{p_z}^2}T} \,
             \ave{\frac{|p_z|}{\omega^2\tilde\omega} \, (m_\pm^2
\theta_\pm')'}';\qquad \tilde\omega\equiv\sqrt{m^2_\pm+p_z^2}
$

\medskip
\noindent Eq.\ (5.56):
$\displaystyle
\ave{\frac{|p_z|}{\omega^2\tilde\omega}} =
            \frac{e^{-x_\pm} - x_\pm E_1(x_\pm )}
                 {2 T^2 x_\pm^2 K_2(x_\pm)},
$

\medskip
\noindent Eq.\ (5.57):
$\displaystyle
S_{H,\rm eff} = -\frac{s}{8}\frac{v_w D_h}{{\aves{v_{p_z}^2}T^3}}  
                   \left(\NBR{e^{-x_\pm} - x_\pm E_1(x_\pm) }
                   \; (m_\pm^2 \theta_\pm')'\right)'.
$

\section{Correction of numerical results}

Due to a programming error, our results for the baryon asymmetry were too
large by several orders of magnitude.  (This error was corrected in the
results presented in ref.\ \cite{CK}.)  The error was discovered in the
course of comparison with our results by the authors of ref.\ \cite{HS}.  
They used the same WKB formalism as we did for deriving the source term in
the diffusion equations, which determines the chiral quark asymmetry that
biases sphalerons to produce the baryon asymmetry.  We are now in
quantitative agreement with them on the size of the baryon asymmetry
produced by charginos in the MSSM.  In addition to correcting this error,
we are also solving the full set of diffusion equations numerically rather
than by using Green's functions.  The latter procedure involved the use of
some approximations which are not necessary in the complete numerical
solution.

Because of the inefficiency of baryogenesis in this model, it is necessary
to assume the CP violating phase Im$(m_2\mu)$ is large, nearly maximal.  
Recent constraints from the electric dipole moments of the electron,
neutron, and especially mercury then imply that the lower-generation
squarks must be quite heavy, on the order of 10 TeV \cite{AKL}.  This kind
of squark spectrum is consistent with what we required for independent
reasons: the chiral quark asymmetry was maximized in this case, and also
the left-handed stop must be this heavy to give sufficiently large
radiative corrections to the light Higgs boson mass, given the need for a
light right-handed to get a strongly first order electroweak phase
transition.

We now present updated figures summarizing the corrected profiles for the
source term and the chiral asymmetry in the bubble wall, and the corrected
dependence of the baryon asymmetry on 
the bubble wall velocity, the wall thickness and the chargino mass
parameters.  We have updated the latter plot to show the exclusion from
the LEP2 limit on the chargino mass, $m_{\chi^\pm} < 104$ GeV$/c^2$.

\FIGURE[t]{
\centerline{\epsfxsize=5.5in \epsfbox{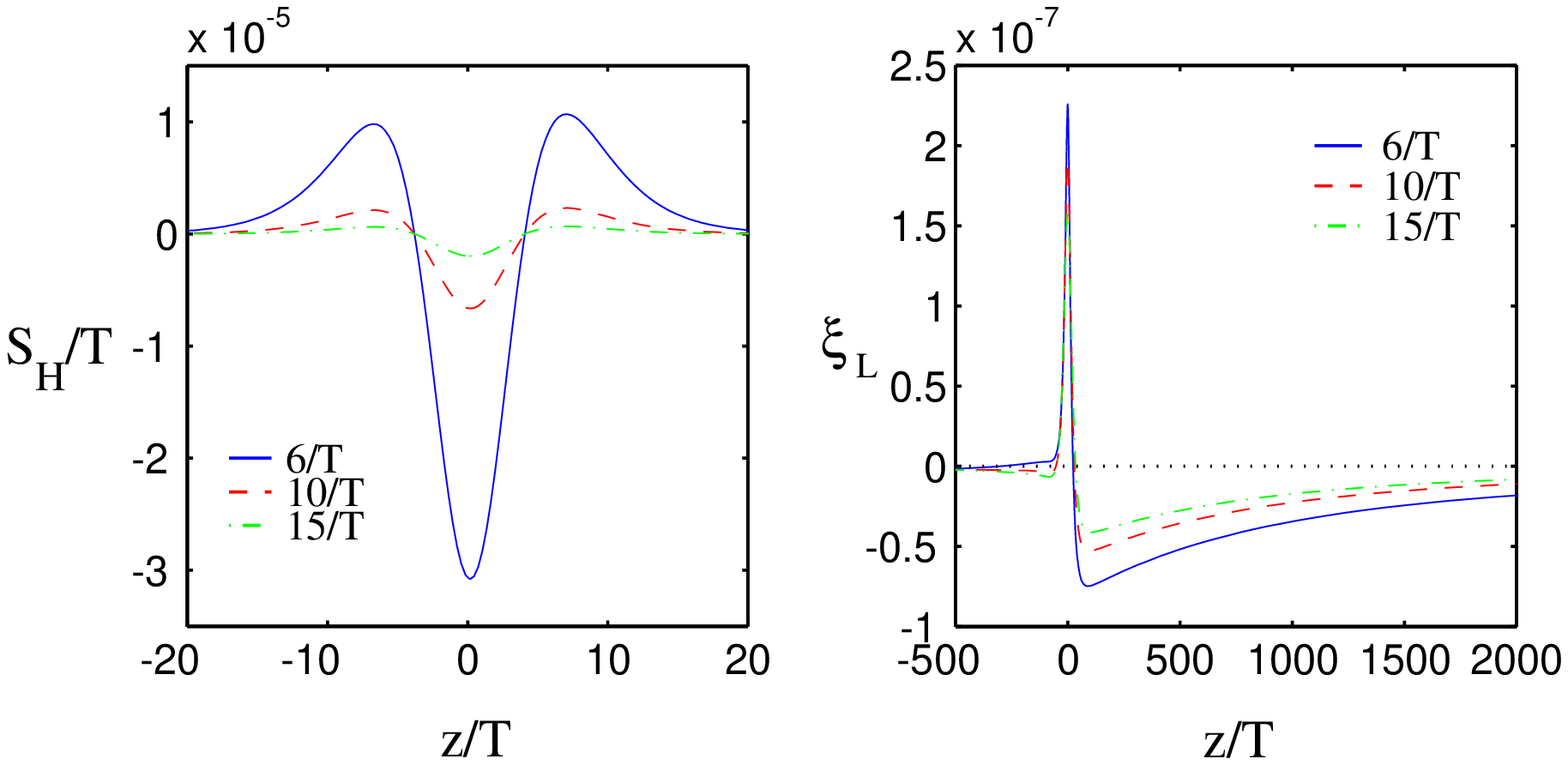}}\\
\vspace{-1.0cm}
\caption{(a) The source for baryogenesis from the chiral classical 
force, eq.\ ({5.57}), for the parameters $\mu = m_2 = 150$ GeV and 
$\ell_w = 6/T$ (solid line), $\ell_w = 10/T$ (dashed line) and
$\ell_w = 15/T$ (dash-dotted line). (b) The left-handed quark 
asymmetry $\xi_{q_L}$, eq.\ (5.44), for the same parameters. The 
distance from the center of the wall $z$, is measured in units 
$1/T$.}
\label{fig1}}

\eject

\FIGURE[h]{
\centerline{\epsfxsize=5.5in \epsfbox{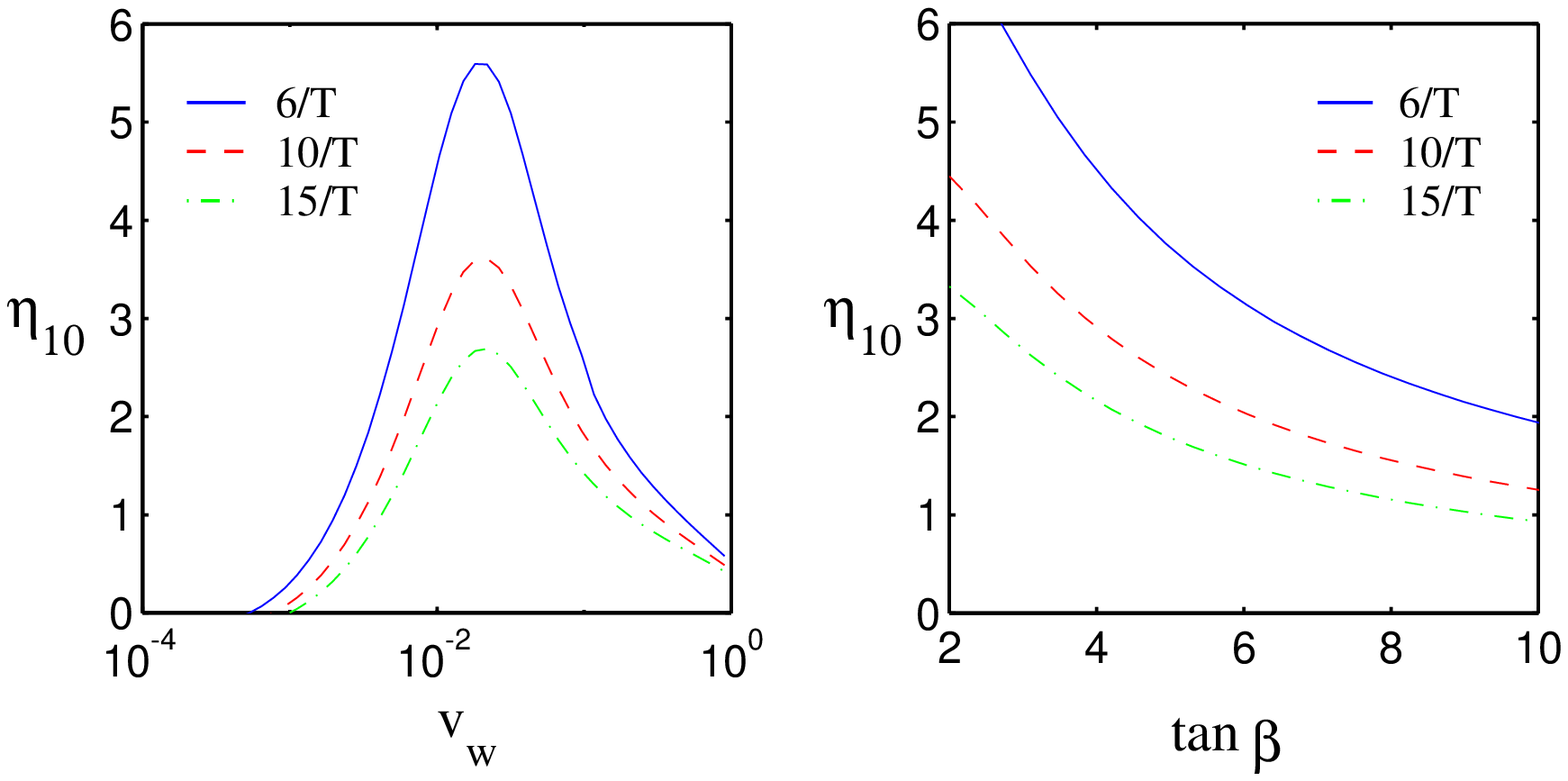}}
\vspace{-1.0cm}
\caption{$\eta_{10}$ for $\mu = m_2 = 150$ GeV and $\sin \delta_\mu=1$ 
(a) as a function of wall velocity and (b) as a function of $\tan \beta$
for a varying wall width, $\ell_w = 6/T$, $10/T$ and $15/T$.} 
\label{fig2}}

\FIGURE[h]{
\centerline{\epsfxsize=5.5in \epsfbox{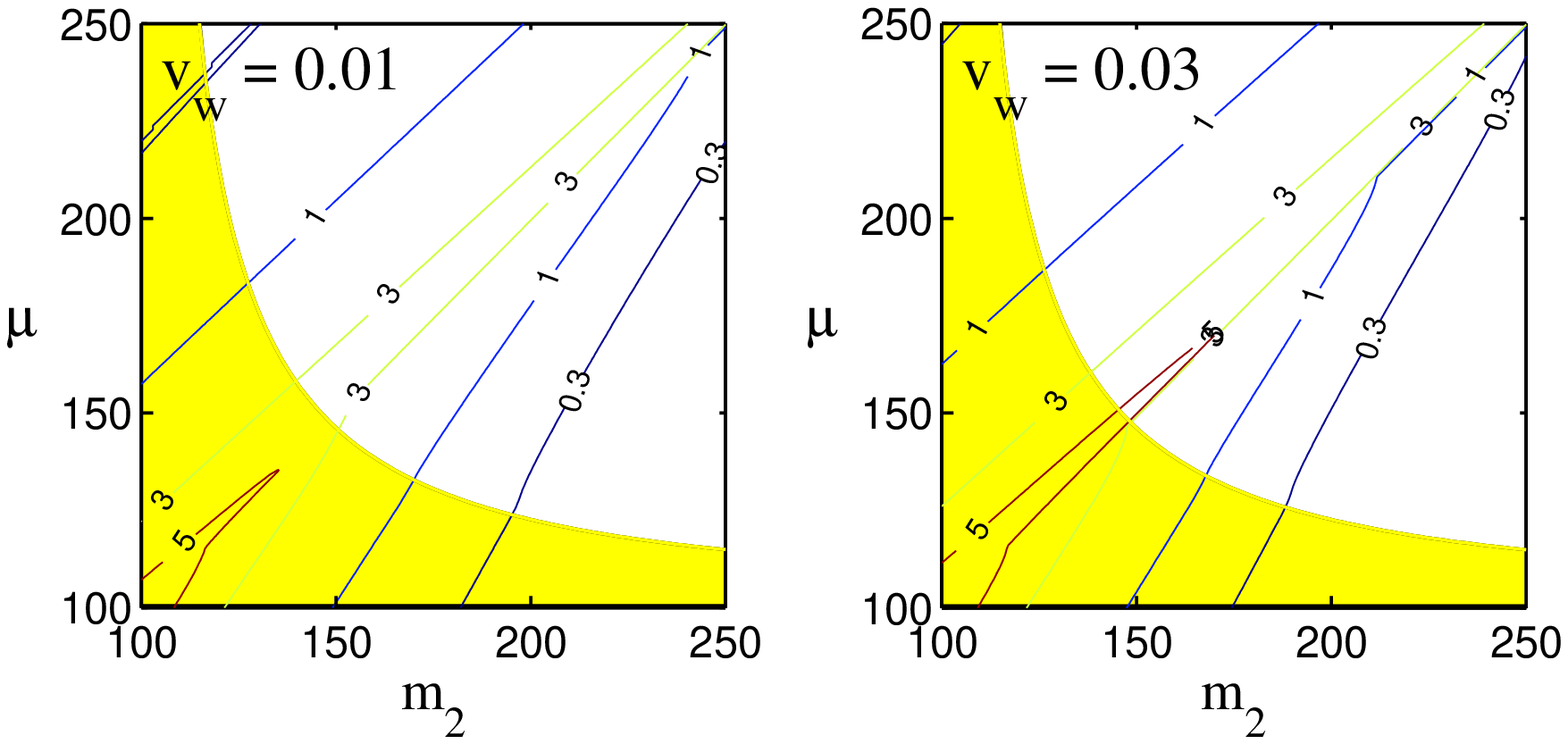}}
\vspace{-1.3cm}
\caption{Contours of constant baryon asymmetry in units $10^{-10}$
with $\sin \delta_\mu = 1$ for (a) $v_w = 0.01$  and (b) $v_w = 0.03$. Mass 
units are GeV$/c^2$. Shaded regions are excluded by the LEP2 limit on the 
chargino mass, $m_{\chi^\pm}>104$ GeV$/c^2$.}   
\label{fig3}}

Our corrected results show that not only must the the phase in the chargino 
mass matrix violate CP nearly maximally to generate a large enough baryon 
asymmetry, but a host of other rather independent parameters must be tuned 
to the optimum as well: we should have $\tan \beta \lsim 3$, the wall 
velocity should be close to the optimal $v_w \simeq 0.02$ and the walls 
should be as narrow as they can come in the MSSM: $\ell_w \simeq 6/T$.

We thank S. Huber for pointing out the discrepancy with our original
results, and T. Prokopec for pointing out the
$\omega^3\to\omega^2\tilde\omega$ correction to
eq.\ (5.54).


\begin{thebibliography}{999}

\bibitem{CK}
J.~M.~Cline and K.~Kainulainen,
Phys.\ Rev.\ Lett.\  {\bf 85}, 5519 (2000)
[hep-ph/0002272].

\bibitem{HS}
S.~J.~Huber and M.~G.~Schmidt,
``Electroweak baryogenesis: Concrete in a SUSY model with a gauge
singlet,''
hep-ph/0003122.

\bibitem{AKL}
S.~Abel, S.~Khalil and O.~Lebedev,
``EDM constraints in supersymmetric theories,''
hep-ph/0103320.

\end{thebibliography}
\end{document}